\documentclass[12pt,letterpaper]{article}
\usepackage[top=1in,left=0.875in,right=0.875in,footskip=1in]{geometry}

\usepackage[utf8]{inputenc}
\usepackage[sort&compress, numbers]{natbib}

\usepackage{nameref,hyperref}

\usepackage[right]{lineno}

\usepackage{array}

\usepackage{microtype}
\DisableLigatures[f]{encoding = *, family = * }

\usepackage[aboveskip=1pt,labelfont=bf,labelsep=period,singlelinecheck=off]{caption}

\usepackage{lastpage,fancyhdr,graphicx}
\usepackage{amsmath}
\usepackage{color}

\definecolor{Gray}{gray}{.25}

\usepackage{booktabs}

\newcommand{\hypertools}{\href{https://github.com/ContextLab/hypertools}{\texttt{HyperTools}}}

\newcounter{example}

\begin{document}
\vspace*{0.1in}

\begin{flushleft}
{\Huge
\textbf\newline{\hypertools: A Python toolbox for visualizing and manipulating high-dimensional data}
}
\newline
\\
{\Large
Andrew C. Heusser\textsuperscript{$\dagger$},
Kirsten Ziman\textsuperscript{$\dagger$},
Lucy L. W. Owen, and
Jeremy R. Manning\textsuperscript{$\ddagger$},
\\
\bigskip
\textbf{Department of Psychological and Brain Sciences, Dartmouth College, Hanover, NH  03755}}
\\
\bigskip
\textsuperscript{$\dagger$} Denotes equal contribution\\
\textsuperscript{$\ddagger$} Address correspondence to \href{mailto:jeremy.r.manning@dartmouth.edu}{jeremy.r.manning@dartmouth.edu}

\end{flushleft}

\begin{abstract}
Data visualizations can reveal trends and patterns that are not otherwise obvious from the raw data or summary statistics.  While visualizing low-dimensional data is relatively straightforward (for example, plotting the change in a variable over time as $(x,y)$ coordinates on a graph), it is not always obvious how to visualize high-dimensional datasets in a similarly intuitive way.  Here we present \hypertools, a Python toolbox for visualizing and manipulating large, high-dimensional datasets.  Our primary approach is to use dimensionality reduction techniques~\cite{Pear01, TippBish99} to embed high-dimensional datasets in a lower-dimensional space, and plot the data using a simple (yet powerful) API with many options for data manipulation (e.g. hyperalignment~\cite{HaxbEtal11}, clustering, normalizing, etc.) and plot styling.  The toolbox is designed around the notion of \textit{data trajectories} and \textit{point clouds}. Just as the position of an object moving through space can be visualized as a 3D trajectory, \hypertools~uses dimensionality reduction algorithms to create similar 2D and 3D trajectories for time series of high-dimensional observations. The trajectories may be plotted as interactive static plots or visualized as animations. These same dimensionality reduction and alignment algorithms can also reveal structure in static datasets (e.g.\ collections of observations or attributes).  We present several examples showcasing how using our toolbox to explore data through trajectories and low-dimensional embeddings can reveal deep insights into datasets across a wide variety of domains.
\end{abstract}


\section*{Introduction}

\begin{quotation}
``\textit{To deal with hyper-planes in a fourteen dimensional space, visualize a 3D space and say `fourteen' to yourself very loudly.  Everyone does it.}'' \\\begin{flushright}\vspace{-0.5cm}--Geoffrey Hinton~\cite{Hint12}\end{flushright}
\end{quotation}

The \hypertools~toolbox is designed to reveal geometric structure in high-dimensional data through visualizations and manipulations.  Modern data visualizations date back to at least the 16\textsuperscript{th} century, when early data pioneers began to develop the sorts of accurate maps and diagrams we might still recognize today~\cite{Frie06, TuftGrav83}.  Visualizations can reveal deep insights and intuitions about geometric structure and patterns in complex datasets by capitalizing on the human visual system's ability to quickly and efficiently extract meaning and structure from highly complex visual information~\cite{UddeEtal16}.  This is perhaps especially true of high-dimensional datasets, where different dimensions or features may interact in complex ways that may not be immediately obvious through conventional summary statistics.

\begin{figure}[tbp]
\centering
\includegraphics[width=1\textwidth]{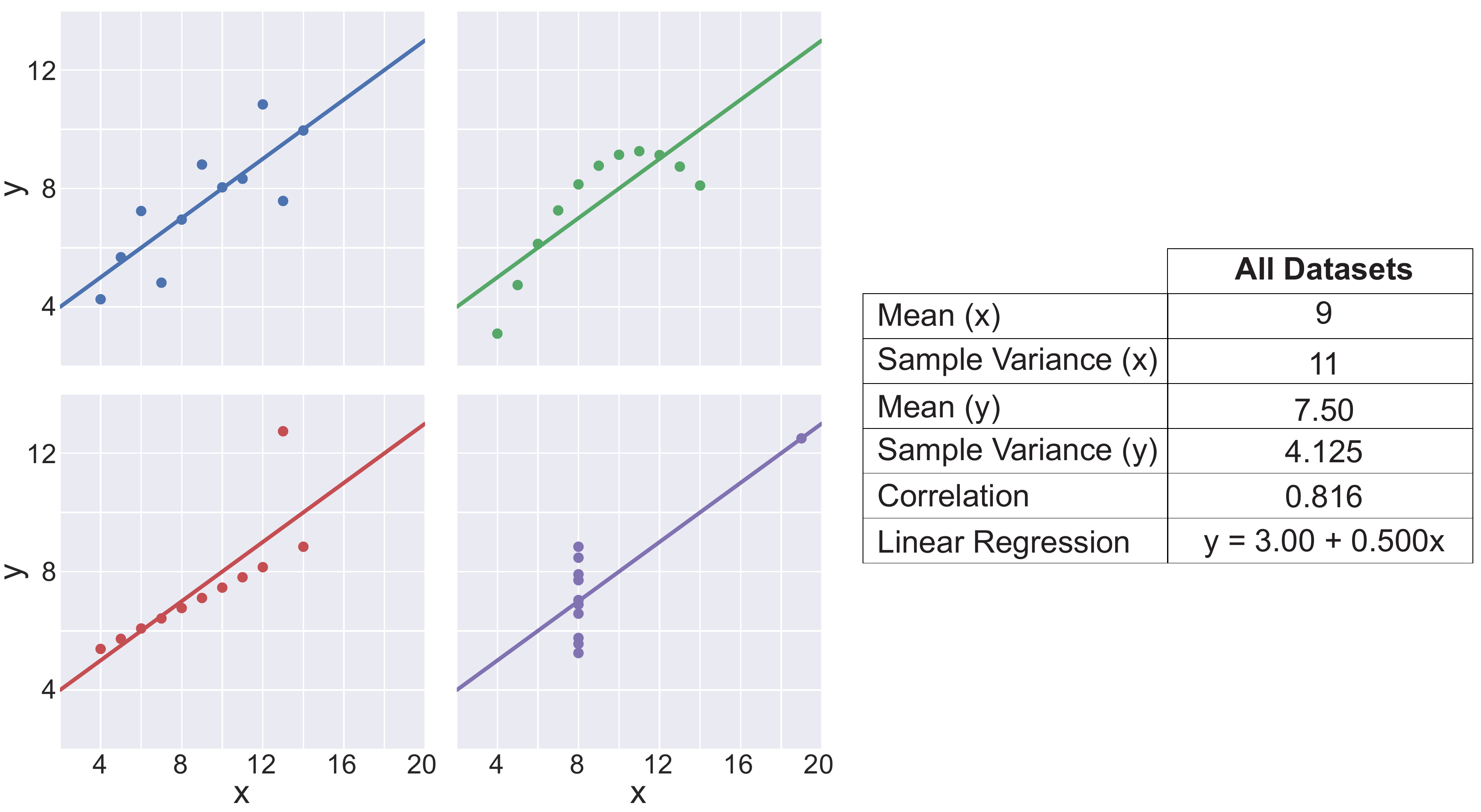}
\caption{\textbf{Anscombe's Quartet.} Each dataset shares the same descriptive statistics but exhibits a unique shape.}
\label{fig:anscomb}
\end{figure}

As an illustration of the potential for summary statistics to mislead in the absence of visualization, consider the classic example, Anscombe's Quartet~\cite{Ansc73} (Fig.~\ref{fig:anscomb}).  Anscombe's Quartet comprises four datasets that share a common statistical profile.  Because the datasets are exactly equal along several common summary measures (mean, variance, trend lines), at first glance they seem highly similar.  However, plotting the datasets and comparing them visually reveals that they differ substantially in structure.
Whereas low-dimensional datasets like those in Anscombe's Quartet can be easily plotted, it is not always obvious how to visualize high-dimensional datasets (e.g.\ with greater than 3 dimensions) in a similarly intuitive way.

\begin{figure}[tbp]
\centering
\includegraphics[width=.75\textwidth]{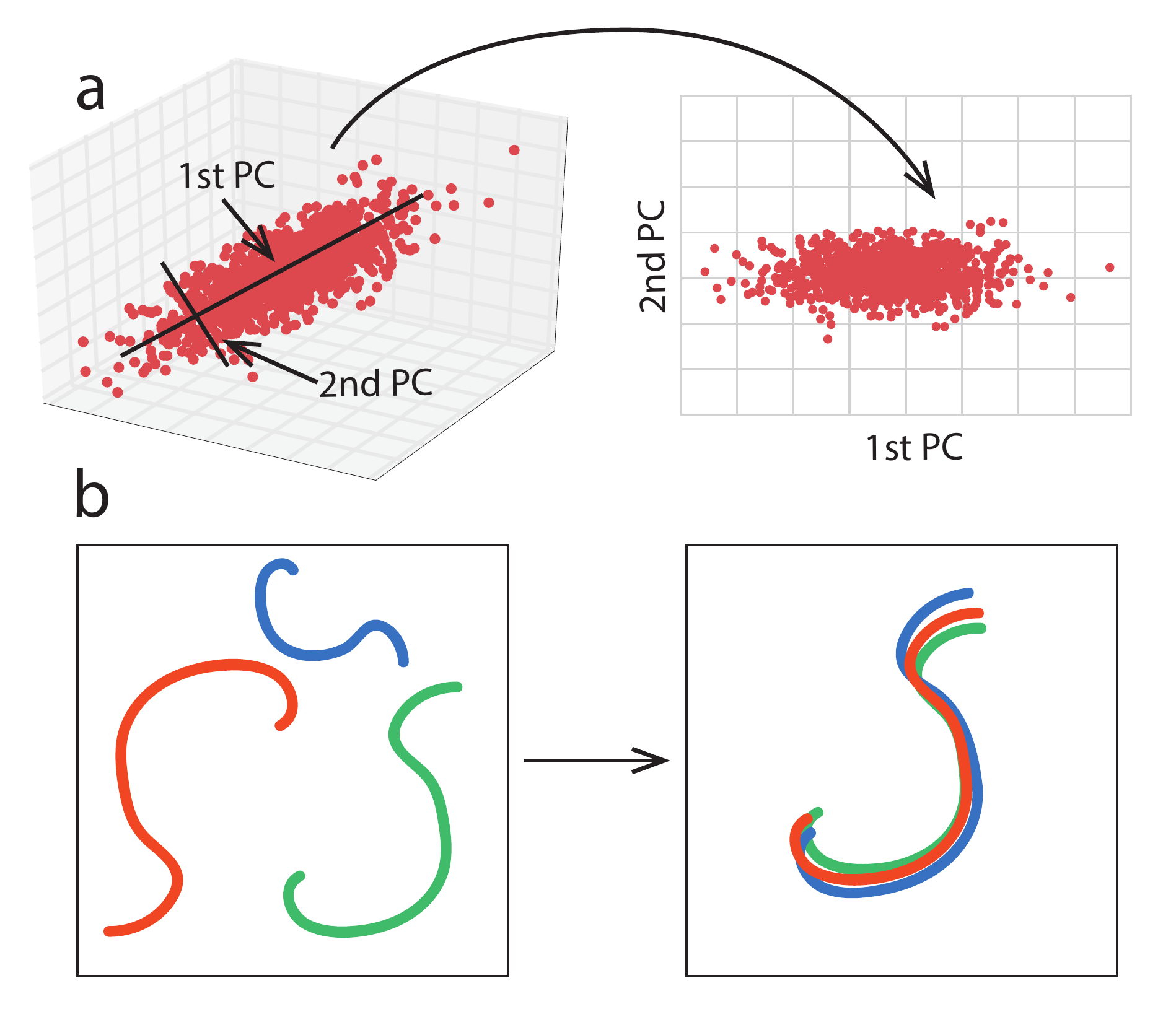}
\caption{\textbf{Visualizing and manipulating high-dimensional data.}  \textbf{a.}  \hypertools~uses dimensionality reduction algorithms to project high-dimensional data onto 2D and 3D plots.  As shown in the panel, the dimensionality reduction algorithm PCA may be used to find the axes that explain the most variance in the original data (left panel).  The data may then be projected onto a small number of those axes to facilitate plotting (right panel).  \textbf{b.} Another important feature of \hypertools~concerns aligning datasets with different fundamental coordinate systems.  The left panel displays three trajectories with similar geometries but different coordinate systems, and the right panel displays how those trajectories may be aligned (via linear transformations) into a common space using hyperalignment.}
\label{fig:methods}
\end{figure}

A number of techniques, collectively referred to as \textit{dimensionality reduction algorithms} have been developed over the past half-century to map high-dimensional data onto lower-dimensional representations that may be more easily manipulated and visualized.  Some well-known examples include Principal Components Analysis (PCA)~\cite{Pear01}, Probabilistic Principal Components Analysis (PPCA)~\cite{TippBish99}, Independent Components Analysis (ICA)~\cite{JuttHera91,
  ComoEtal91}, Multidimensional Scaling (MDS)~\cite{Torg58}, and $t$-Distributed Stochastic Neighbor Embedding (t-SNE)~\cite{MaatHint08}.  While the details of these algorithms differ, they each provide a means of obtaining a low-dimensional representation of the original high-dimensional dataset that preserves many of the geometric properties (e.g.\ the overall covariance structure of the data, data grouping, etc.) to the extent possible within a low-dimensional space.  In the \hypertools~toolbox, we leverage these dimensionality reduction algorithms (Fig.~\ref{fig:methods}a) to aide in visualizing high-dimensional data.

A second class of algorithms leveraged in our toolbox provide techniques for manipulating and aligning different high-dimensional datasets (Fig.~\ref{fig:methods}b).  These algorithms draw inspiration from the Procrustean transformation~\cite{Scho66}, which computes the affine transformations (i.e.\ translation, reflection, rotation, and scaling) that bring one trajectory into alignment with another (in terms of minimizing the mean squared Euclidean distances between the corresponding points).  The hyperalignment algorithm~\cite{HaxbEtal11} and the Shared Response Model (SRM)~\cite{ChenEtal15} extend this technique to find a common set of transformations that bring many (more than two) high-dimensional trajectories into common alignment.  Our \hypertools~toolbox leverages these alignment algorithms to allow users to manipulate and compare high-dimensional data, even when the dimensions (features) of original observations are different (e.g.\ brain patterns from different people, observations from different modalities, etc.).

Taken together, the \hypertools~toolbox provides a set of powerful functions for visualizing and manipulating high-dimensional data using dimensionality reduction and data alignment algorithms.  The toolbox is designed with ease of use as a primary goal, such that complex visualizations and analyses may often be carried out with a single line of code.  Another major goal is to enable users to easily produce visually appealing publication-quality plots, also often with only a single line of code.  Our toolbox is open-source and is distributed with the MIT License.

In the next section we provide a detailed overview of the components of the \hypertools~toolbox and describe how the codebase is organized.  We then describe a series of analyses of datasets from a wide array of domains to highlight many of the main functions of the toolbox.

\section*{Materials and Methods}
\subsection*{Overview}
The \hypertools~toolbox is written in Python and can be downloaded from our \href{https://github.com/ContextLab/hypertools}{GitHub} page or with \href{https://pip.pypa.io/en/stable/installing/}{\texttt{pip}}:
\begin{align}
\texttt{pip install hypertools}
\end{align}

\hypertools~depends on the following open-source software packages: \href{http://matplotlib.org/}{\texttt{Matplotlib}}~\cite{Hunt07} for plotting functionality, \href{http://seaborn.pydata.org/}{\texttt{Seaborn}}~\cite{WaskEtal16} for plot styling,  \href{https://github.com/scikit\-learn/scikit\-learn}{\texttt{scikit-learn}}~\cite{PedrEtal11} for data manipulation (dimensionality reduction, clustering, etc.), and \href{https://github.com/allentran/pca-magic}{\texttt{PPCA}} for inferring missing data using PPCA.  The toolbox also includes a port of the hyperalignment algorithm~\cite{HaxbEtal11} from the \href{https://github.com/PyMVPA/PyMVPA}{\texttt{PyMVPA}} library, as well as the shared response model from the \href{https://github.com/IntelPNI/brainiak}{\texttt{BrainIAK}} toolbox, as an alternative data alignment technique.  In addition to providing a simple interface to several functions from these libraries, \hypertools~adds a number of custom arguments to facilitate data visualization and manipulation of high-dimensional data. Table~\ref{tab:organization} provides summary of the \hypertools~code base.  In the remainder of this section, we provide descriptions of the primary toolbox functions, but we have not provided an exhaustive list.  A feature-complete description of the API may be found on the project's \href{https://github.com/ContextLab/hypertools}{GitHub} page and in the documentation included with the toolbox download.

\begin{table}[tbp]

\centering
    \begin{tabular}{m{2in} m{4.4in}}
    \toprule
    \textbf{Filename} & \textbf{Description} \\ \hline
    \texttt{plot/plot.py} & Main plotting function: parses arguments, dispatches to \\
    & \texttt{static.py} and \texttt{animate.py} \\ 
    \texttt{plot/static.py} & Handles all static plot logic \\
    \texttt{plot/animate.py} & Handles all animated plot logic \\ 
    \texttt{tools/align.py} & Aligns the coordinate space of a list of matrices using \\
    & hyperalignment\\
	\texttt{tools/cluster.py} & Parcellates observations into discrete clusters using $k$-means\\
    & clustering\\ 
    \texttt{tools/describe\_pca.py} & Analyzes and plots how many principal components are needed\\
    & to capture the covariance structure of the data\\
    \texttt{tools/missing\_inds.py} & Find \texttt{nan}s in data and returns indices\\
    \texttt{tools/normalize.py} & $z$-scores rows/columns of matrices\\
    \texttt{tools/df2mat.py} & Converts \texttt{Pandas} dataframes to \texttt{Numpy} arrays \\ 
    \texttt{tools/procrustes.py} & Aligns the coordinate spaces of two arrays \\ 
    \texttt{tools/reduce.py} & Reduces the dimensionality of one or more arrays using PCA\\
    & and PPCA\\ 
    \texttt{\_externals/srm.py} & Implements the Shared Response Model (alternative alignment\\
    & algorithm) \\ 
    \texttt{\_shared/helpers.py} & Collection of helper functions used across many files\\
	\bottomrule
    \end{tabular}
\vspace{0.1in}
\caption{\textbf{\hypertools~code organization.}  The table lists the main files and functions that comprise the toolbox.  We provide a feature complete description of the API on the project's \href{https://github.com/ContextLab/hypertools}{GitHub} page and in the documentation included with the toolbox download.}
\label{tab:organization}
\end{table}

Nearly all of the \hypertools~functions may be accessed through the main \texttt{plot} function.  This design enables complex data analysis, data manipulation, and plotting to be carried out in a single function call.  There are two general types of plots supported by the toolbox: \textit{static plots} and \textit{animated plots}.

\subsection*{Static Plots}
Accessing the \hypertools~plot functionality entails first loading the to-be-analyzed dataset into the Python workspace and converting it to a \texttt{Numpy} array~\cite{WaltEtal11} or a \texttt{Pandas} dataframe~\cite{Mcki10}. The format of the data should be samples ($S$) by features ($F$). Once the dataset conforms to this format, simply import the library and call the plot function:
\begin{align}
& \texttt{import hypertools as hyp} \\
& \texttt{hyp.plot(data)}
\end{align}

By default (i.e.\ with no additional arguments specified), this function will perform dimensionality reduction (using PCA), convert the $S \times F$ \texttt{data} matrix into an $S \times 3$ matrix, and then create an interactive 3D line plot that can be explored by using the mouse to rotate the plot. If there are \texttt{nan}s present in the dataset, these missing values will be automatically interpolated using PPCA~\cite{TippBish99}. If $F < 3$, a 2D plot is created instead of a 3D plot.  This simple interface to plotting is deceptively powerful: with a single command, the toolbox automatically fills in missing data and determines whether to create a 2D or 3D plot (reducing the dimensionality of the observations as needed).

\hypertools~can also accommodate lists of \texttt{Numpy} arrays or \texttt{Pandas} dataframes (only single-level indexed dataframes are currently supported):
\begin{align}
\texttt{hyp.plot([array1, array2, array3])}
\end{align}

When passed a list of arrays, \hypertools~will plot each array in a distinct color.  Colors and styling can be customized in a several ways.  Like \texttt{Matplotlib}, \hypertools~can parse format strings passed as positional arguments.  For example:
\begin{align}
& \texttt{hyp.plot(array1, 'k-')} \\
& \texttt{hyp.plot([array1, array2, array3], ['bo', 'r--', 'g:'])}
\end{align}

Line colors may also be specified via the \texttt{color} (or \texttt{colors}) keyword argument:
\begin{align}
& \texttt{hyp.plot(array1, color='g')} \\
& \texttt{hyp.plot([array1, array2, array3], colors=['b', '\#FF0000', (.3, .5, .4)])}
\end{align}
Colors may be defined using format strings, hex codes, RGB values, or a mix of these formats. Rather than specifying the colors of each data array, colors may instead be specified for each individual sample by providing labels for each sample:
\begin{align}
\texttt{hyp.plot(data, group=group\_labels)},
\end{align}
where \texttt{group\_labels} is a list of length $S$ (number of samples).  (Lists of group label arrays are also supported, e.g.\ if the data are passed in as a list of arrays; the list of labels must be of the same length as the list of data arrays.)

\hypertools~parses this function call by sub-dividing each data matrix into new lists defined by each unique label in \texttt{group\_labels}.  For example, if each sample label is a string from the set \texttt{(`a', `b', `c')} then each of these unique labels will be assigned a unique color, and the datapoints assigned to each label will be assigned that label's color.

In addition to specifying string labels for each sample, \hypertools~also supports numerical labeling.  If \texttt{group\_labels} is a list of numbers, \hypertools~will bin the range covered by those numerical values (excluding \texttt{nan}s, \texttt{None}s, and \texttt{inf}s) into $n$ evenly spaced bins (default: $n = 100$) and map these values onto a color palette.  The color palette used for this mapping may also be customized using the \texttt{palette} keyword:
\begin{align}
\texttt{hyp.plot([array1, array2], palette='husl')}
\end{align}
All \texttt{Matplotlib} and \texttt{Seaborn} color palettes and plot styles are supported by \hypertools.

In addition to specifying group-level labels (which are used to determine the colors of each sample), each sample may also be labeled with an additional text label that may be shown (as text) on the plot.  The \texttt{label} and \texttt{labels} keyword arguments allow the user to define a list of strings (or a list of lists of strings) to be displayed next to each sample datapoint with an arrow pointing to it.  Each list of labels must be of length $S$ (number of samples).  (The \texttt{None} value may be used to specify ``blank'' labels, which will not show up on the plot.)

By default, all datapoint labels are shown if the \texttt{label} or \texttt{labels} keyword is specified.  However, \hypertools~also supports a ``data exploration'' mode whereby the datapoint labels will only be shown when the mouse pointer hovers over the corresponding datapoint:
\begin{align}
\texttt{hyp.plot(data, labels=['a', None, 'a', 'b'], explore=True)}
\end{align}
This plotting mode is especially useful when there are many datapoints, or when the data labels are long.  If \texttt{explore} is set to \texttt{True} and no labels are specified, the labels will be auto-generated as an index and the PCA coordinate (e.g.\ \texttt{'45: (3.0, 4.0, 5.0)'}). Note: at the time of this writing, the \texttt{labels} and \texttt{explore} arguments are only supported for 3D static plots.

\subsection*{Animated Plots}
\begin{figure}[tbp]
\centering
\includegraphics[width=\textwidth]{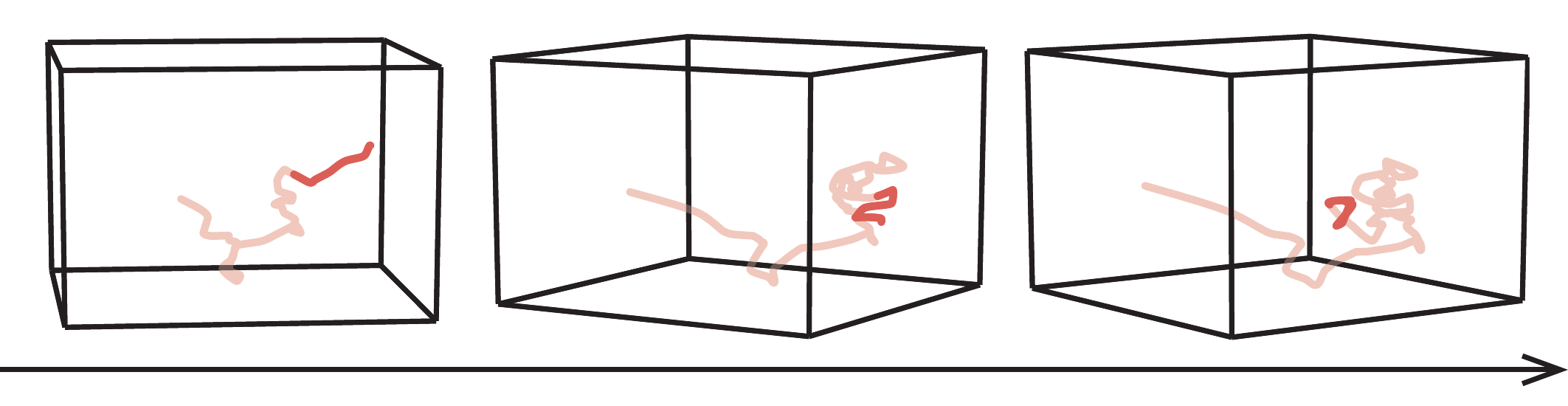}
\caption{\textbf{Frames from an animated plot.}  Three frames (with time increasing moving from left to right) from an example animation are displayed in each panel.}
\label{fig:animation}
\end{figure}

Animated 3D plots are especially useful for visualizing high-dimensional timeseries data.  To create an animation, simply toggle the \texttt{animate} keyword:
\begin{align}
\texttt{hyp.plot(data, animate=True)}
\end{align}
This will create a 3D animated representation of the data, where the animation occurs over the rows of the \texttt{data} matrix.  As with static plots, the user may pass a list of data matrices to plot multiple datasets on a single plot, and format strings and keyword arguments may be used to customize the plot appearance.  Each frame of the animation displays a portion of the total data trajectory enclosed in a cube (Fig.~\ref{fig:animation}).  In successive frames, the displayed portion of the data trajectory advances by a small amount, and the camera angle rotates around the cube, providing visual access to different aspects of the data as the animation progresses.

The formats of animated \hypertools~plots may be customized using the following keyword arguments: \texttt{duration} specifies the animation duration in seconds, \texttt{tail\_duration} specifies the duration of the trailing tail in seconds, \texttt{rotations} specifies the number of camera rotations around the data (over the course of the entire animation), \texttt{zoom} will zoom the camera in (positive number) or out (negative number) from the data, and setting \texttt{chemtrails=True} will plot a low opacity version of samples prior to the currently active trajectory so that the full structure and history of the data may be visualized.   For a complete list of animation-specific arguments, please see the API documentation.  Both animated and static plots can be saved by including the \texttt{save\_path} argument (with the file extension included):
\begin{align}
& \texttt{hyp.plot(data, save\_path='path/to/the/file.pdf')}\label{eqn:static}\\
& \texttt{hyp.plot(data, animate=True, save\_path='path/to/the/file.mp4')}\label{eqn:animated},
\end{align}
where Snippet~\ref{eqn:static} saves a static plot to a resolution-independent PDF, and Snippet~\ref{eqn:animated} saves an animation as an MP4 movie.  Note that saving animated plots requires that \texttt{ffmpeg} is installed on your computer; see the  API documentation for more details.

\subsection*{Reduce}
When passed high-dimensional data, the \texttt{plot} function uses PPCA to fill in missing data and PCA to project the data onto 3 dimensions.  We provide API access to the \texttt{reduce} function that underlies these transformations.  At its core, the \texttt{reduce} function is primarily a wrapper for \texttt{scikit-learn}'s PCA function and the \texttt{PPCA} package.  \hypertools~extends the functionality of these tools by providing an easier-to-use syntax and adding support for lists of matrices.  The function may be used as follows:
\begin{align}
\texttt{reduced\_data = hyp.tools.reduce(data, ndims=3)}
\end{align}

\begin{figure}[tbp]
\centering
\includegraphics[width=0.6\textwidth]{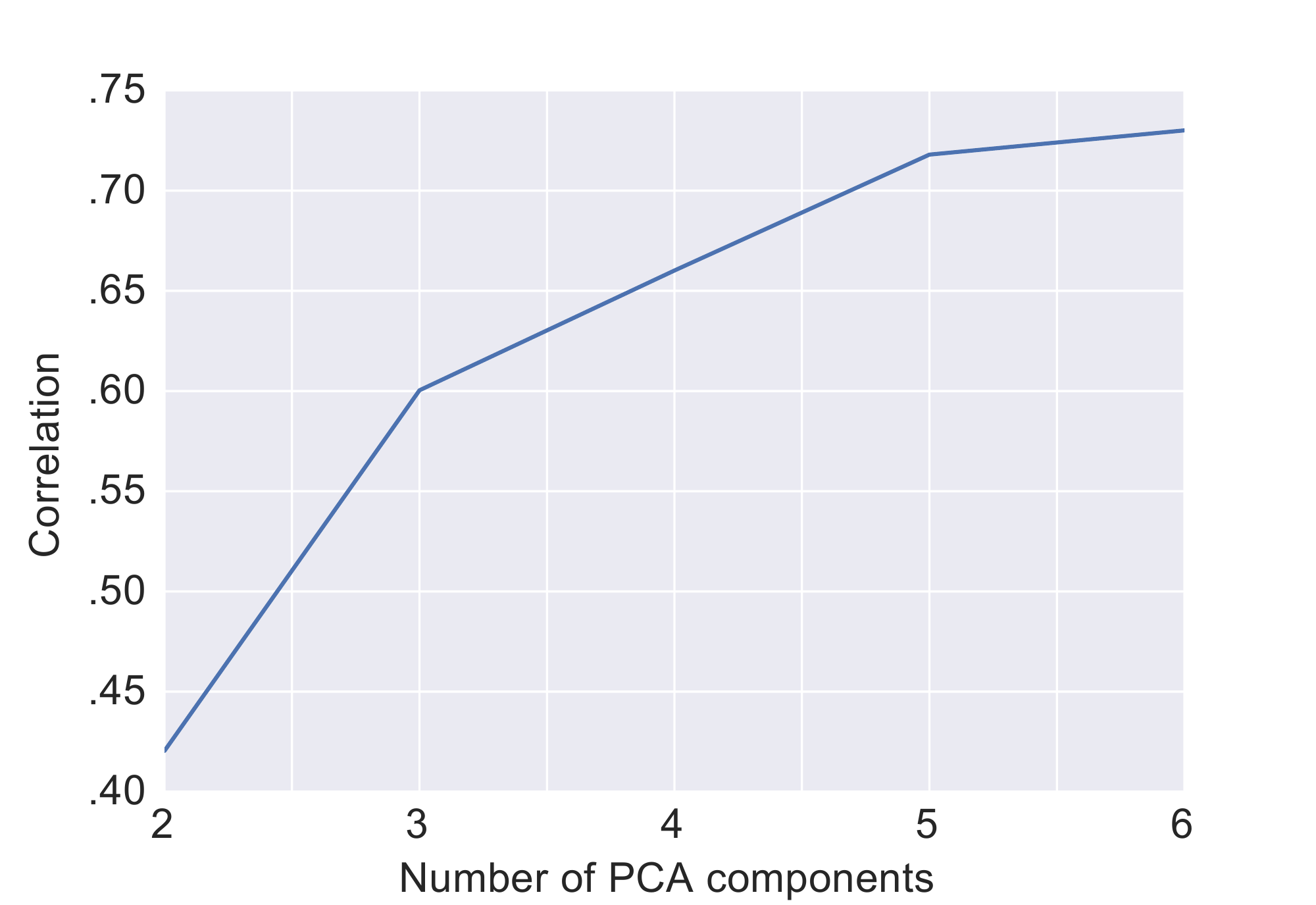}
\caption{\textbf{Covariance preserved as a function of the number of principal components.}  For an example dataset (Example 2, \textit{Results}) the panel displays the correlation between the upper triangle of the across-sample covariance matrices of the reduced versus original data, as a function of the number of principal components.}
\label{fig:describe}
\end{figure}

Because dimensionality reduction results in information loss (relative to the original dataset), it is important to consider how accurately a low-dimensional projection of the data reflects the original high-dimensional dataset.  \hypertools~includes a function that plots the correlation between the covariance matrices of the reduced and full datasets, as function of the number of principal components in the reduced dataset (Fig.~\ref{fig:describe}):
\begin{align}
\texttt{fig, ax, data = hyp.tools.describe\_pca(data)}
\end{align}

The \texttt{describe\_pca} function computes these correlations iteratively (i.e.\ starting with one principal component, then two, then three, etc.) until a local maximum is detected.  The resulting plot provides insights into the increase in explanatory power (in terms of the across-sample covariance) associated with each new principal component.

\subsection*{Align}
Two or more datasets may share geometrical structure, but reside in different coordinate systems. Hyperalignment is a method that aligns the representational spaces over a list of datasets, effectively co-registering them to a common space~\cite{HaxbEtal11}.  Using linear transformations, hyperalignment find a common space that minimizes the distance between two or more datasets (Fig.~\ref{fig:methods}b).  Aligning them to a common space allows one to visualize commonalities between the two different kinds of data.  The \texttt{align} function accepts a list of arrays as input and returns a hyperaligned list of arrays in a common geometric space:
\begin{align}
\texttt{hyperaligned\_list = hyp.tools.align([array1, array2, array3])}
\end{align}

In addition to supporting alignment via the hyperalignment algorithm proposed by~\cite{HaxbEtal11}, we have also added support for alignment via the Shared Response Model~\cite{ChenEtal15}, which was ported from the \texttt{BrainIAK} toolbox:
\begin{align}
\texttt{SRM\_aligned\_list = hyp.tools.align([array1, array2, array3], method='SRM')}
\end{align}

\subsection*{Cluster}
Some datasets exhibit \textit{clustering} tendencies, whereby the data may be divided into discrete groups of similar or related samples (i.e.\ samples that are comprised of similar features).  When these discrete groups are unlabeled or unknown, clustering algorithms provide heuristics for recovering these clusters of similar samples automatically.  \hypertools~incorporates the $k$-means clustering algorithm~\cite{HartWong79} to facilitate automatic data clustering.  Given a pre-chosen number of clusters, $k$, the \texttt{cluster} keyword argument to the \texttt{plot} function uses $k$-means clustering to automatically assign each observation to a cluster, and then colors each observation's point according to its cluster membership:
\begin{align}
\texttt{hyp.plot(data, n\_clusters=k)}
\end{align}

We also expose the $k$-means clustering algorithm directly through the \texttt{cluster} function:
\begin{align}
\texttt{cluster\_labels = hyp.tools.cluster(data, n\_clusters=k)}
\end{align}
The \texttt{cluster} function wraps the \texttt{scikit-learn} implementation of $k$-means clustering and extends it to work with lists of data matrices.

\section*{Results}
\subsection*{Example \stepcounter{example}\arabic{example}: Visualizing hypercubes in 3D}

To illustrate how a user might visualize high-dimensional data with \hypertools, we start by examining four synthetic datasets with unique, known structures. We generated datasets of one cube (3 dimensions) and three hypercubes of increasing dimensionality (4, 5 and 6 dimensions), each comprised of 100 points along each of their respective edges.  We then used \hypertools~to project the hypercubes into 3 dimensional space (using PCA) and visualize the result. 

\begin{figure}[tbp]
\centering
\includegraphics[width=.75\textwidth]{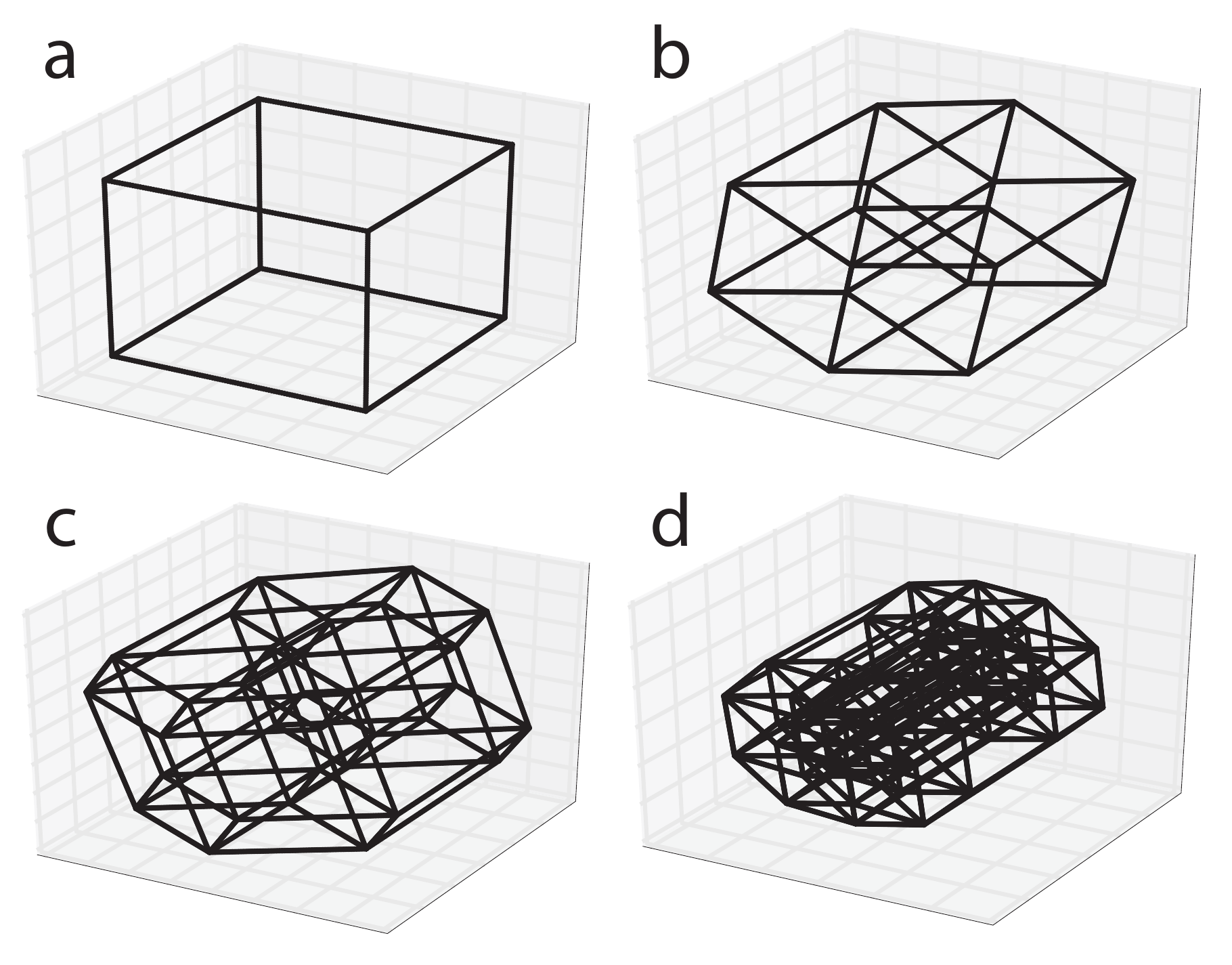}
\caption{\textbf{Hypercubes with increasing dimensionality.} Each dataset comprises 100 evenly spaced points along each edge of the corresponding cube with dimensionality \textbf{a.} 3, \textbf{b.} 4, \textbf{c.} 5, and \textbf{d.} 6.}
\label{fig:hypercubes}
\end{figure}

Figure~\ref{fig:hypercubes} illustrates how projecting hypercubes of different dimensionalities into 3 dimensional space distorts some aspects of their shapes, while preserving others. In the original (high-dimensional) data, all edges of each respective cube are of equal length, and each vertex comprises $n$ adjacent edges converging orthogonally (where $n$ is the dimensionality of the hypercube). However, in Fig.~\ref{fig:hypercubes}, some edges appear longer than others, and some vertices appear to form acute and obtuse angles.

Despite these differences, many of the underlying structural components are accurately reflected in the visualization. Namely, the visualization of each $n$-dimensional cube correctly depict $2^n$ vertices, $2^{n-1}*n$ edges, and $n$ edges converging at each vertex. Each edge is also reliably reconstructed as a straight (rather than curved) line segment.  The visualizations also depict increasing complexity with increasing dimensionality.

\subsection*{Example \stepcounter{example}\arabic{example}: Dimensionality reduction and clustering with various types of mushrooms}

In this section, we highlight the dimensionality reduction and clustering capabilities of \hypertools. We retrieved the `mushroom classification' dataset from the  \href{https://www.kaggle.com/uciml/mushroom-classification}{Kaggle} database. The dataset contains annotated descriptive features of 8,124 mushrooms spanning 23 mushroom species from the \textit{Audubon Society Field Guide to North American Mushrooms}~\cite{Linc81}. Each observation comprises a list of 22 descriptive features (e.g.\ cap shape, cap surface, habitat, etc.) along with a tag identifying each mushroom exemplar as poisonous or non-poisonous (features for five example mushrooms are shown in Tab.~\ref{tab:mushrooms}).

\begin{table}[tbp]
\centering
\begin{tabular}{lllllllll}
\toprule
{} & class & cap-shape & cap-surface & cap-color & bruises & odor &  ... & habitat \\
\midrule
0 &     p &         x &           s &         n &       t &    p &  ... &       u \\
1 &     e &         x &           s &         y &       t &    a &  ... &       g \\
2 &     e &         b &           s &         w &       t &    l &  ... &       m \\
3 &     p &         x &           y &         w &       t &    p &  ... &       u \\
4 &     e &         x &           s &         g &       f &    n &  ... &       g \\
\bottomrule
\end{tabular}
\vspace{0.1in}
\caption{\textbf{Example of mushrooms dataset.} The dataset contains annotated features (columns) of each mushroom (row), along with labels indicating whether each mushroom is poisonous or non-poisonous (not shown).}
\label{tab:mushrooms}
\end{table}

\begin{figure}[tbp]
\centering
\includegraphics[width=.75\textwidth]{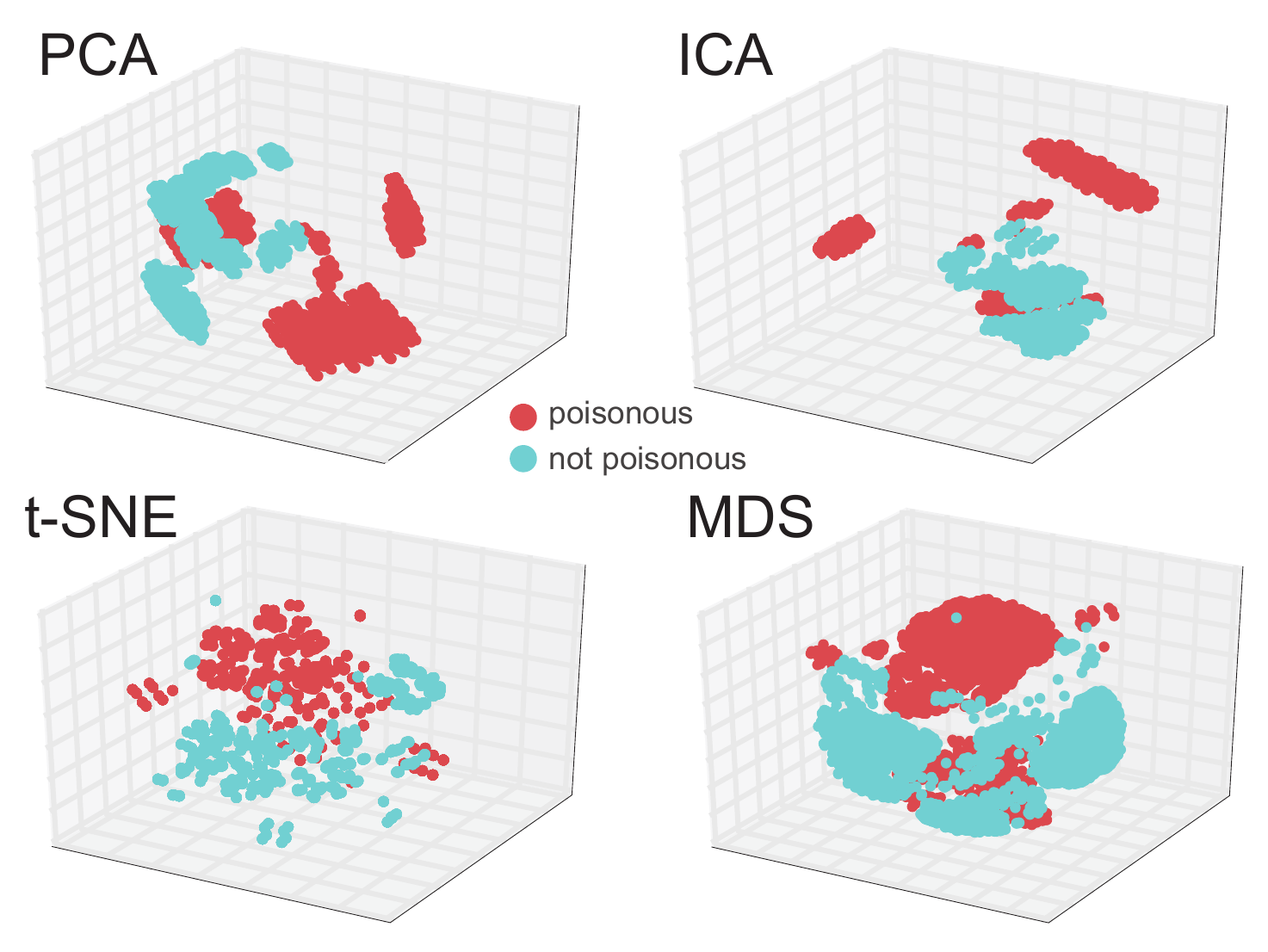}
\caption{\textbf{Three-dimensional embeddings of the mushrooms dataset using several dimensionality reduction techniques.} Each point represents a sample (mushroom).  Red dots indicate poisonous mushrooms and blue indicate non-poisonous mushrooms.} \label{fig:mushproj}
\end{figure}

Because the mushroom features are provided as character strings, they must be transformed into numerical vectors to plot them.  When passed a \texttt{Pandas} dataframe with columns containing text, \hypertools~automatically converts the data into a binary matrix, where each column reflects one of the unique values of one of the features.  The underlying function for converting dataframes into matrices may also be called directly:
\begin{align}
\mathtt{matrix = hyp.tools.df2mat(dataframe)}
\end{align}

Plotting the resulting matrix with \hypertools~reveals a striking clustered structure.  Overall, the samples appear to cluster by whether or not they are poisonous, but they also appear to group into sub-clusters (Fig.~\ref{fig:mushproj}). By default, \hypertools~uses PCA for dimensionality reduction, but different dimensionality reduction techniques can reveal distinct geometrical properties of a dataset. To highlight this, we plotted the transformed binary matrix using several dimensionality reduction techniques (PCA, ICA, $t$-SNE, and MDS) to visualize their effects on clustering (Fig.~\ref{fig:mushproj}). Each technique produces a unique low-dimensional projection of the data, highlighting distinct structural aspects.

\begin{figure}[tbp]
\centering
\includegraphics[width=0.5\textwidth]{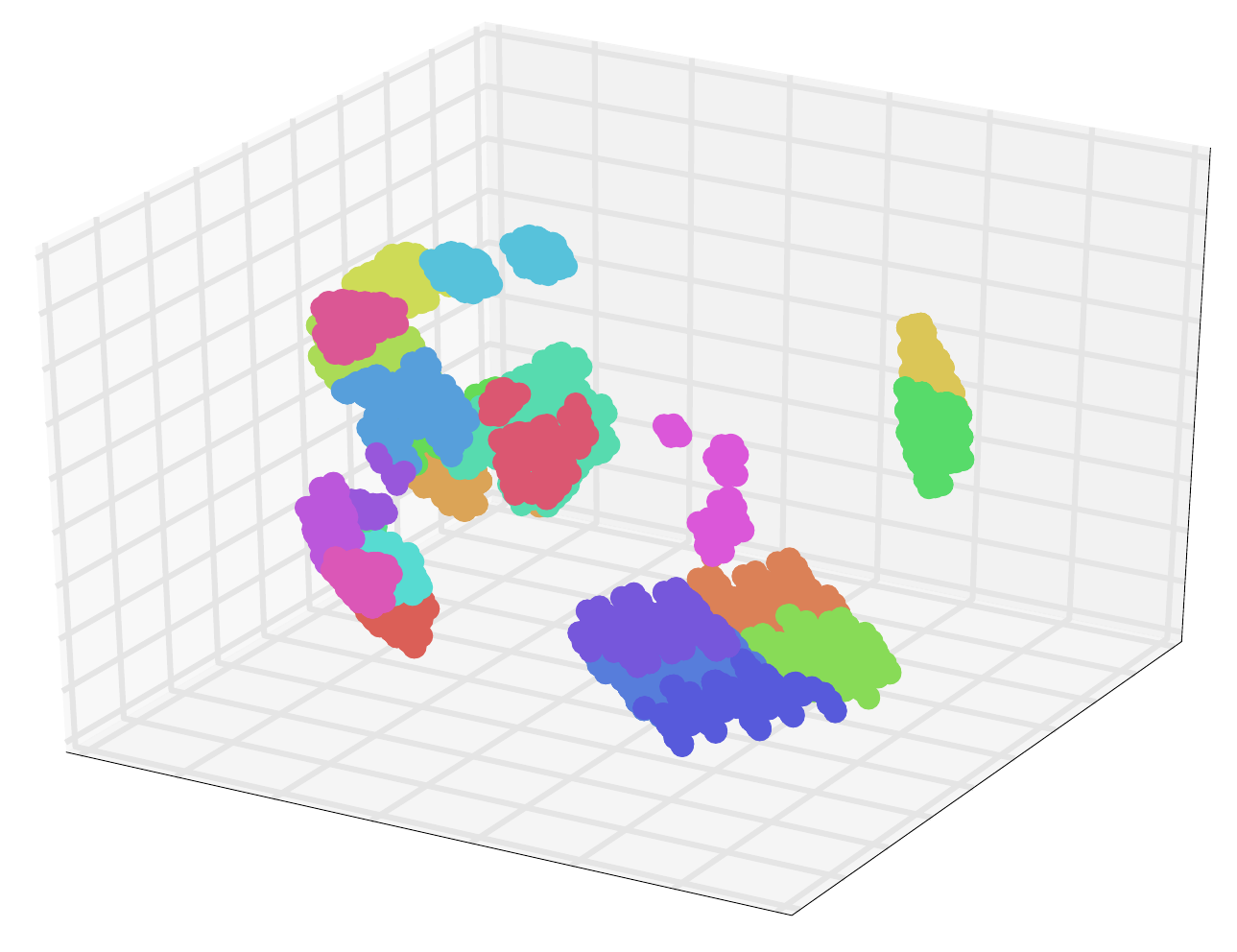}
\caption{\textbf{Mushrooms dataset, colored by $k$-means cluster.}} \label{fig:mushclust}
\end{figure}

To highlight the sub-clustering structure in this dataset,  we use the \texttt{n\_clusters} argument to \texttt{plot}:
\begin{align}
\mathtt{hyp.plot(mushrooms\_data, n\_clusters=23)}
\end{align}
This command relies on $k$-means clustering to empirically derive cluster labels, and then plot each cluster in a different color (Fig.~\ref{fig:mushclust}).

\subsection*{Example \stepcounter{example}\arabic{example}: Exploring factors that influence educational outcomes.}
Next, we analyzed an education dataset containing, for each of 480 students around the world, performance ratings (high, medium, and low performance), demographic descriptors (e.g.\ gender, nationality, place of birth, etc.) as well as classroom behaviors (number of times the student raised their hand, days absent from class, number of times the student visited online resources, etc.) and others (features for five example students are displayed in Tab.~\ref{tab:education}; for full list of features and to download the data, see the \href{https://www.kaggle.com/aljarah/xAPI-Edu-Data}{Kaggle database}).  Given a dataframe with the student features, \hypertools~automatically converts this into a binary data matrix (as described above) for visualization.

\begin{table}[tbp]
\centering
\begin{tabular}{llllllll}
\toprule
{} & gender & NationalITy & PlaceofBirth &     StageID & GradeID &  ... & Class \\
\midrule
0 &      M &          KW &       KuwaIT &  lowerlevel &    G-04 &  ... &     M \\
1 &      M &          KW &       KuwaIT &  lowerlevel &    G-04 &  ... &     M \\
2 &      M &          KW &       KuwaIT &  lowerlevel &    G-04 &  ... &     L \\
3 &      M &          KW &       KuwaIT &  lowerlevel &    G-04 &  ... &     L \\
4 &      M &          KW &       KuwaIT &  lowerlevel &    G-04 &  ... &     M \\
\bottomrule
\end{tabular}
\vspace{0.1in}
\caption{\textbf{Example features in education dataset.} The dataset contained categorical and numerical features, as well as student performance labels.}
\label{tab:education}
\end{table}

\begin{figure}[tbp]
\centering
\includegraphics[width=1\textwidth]{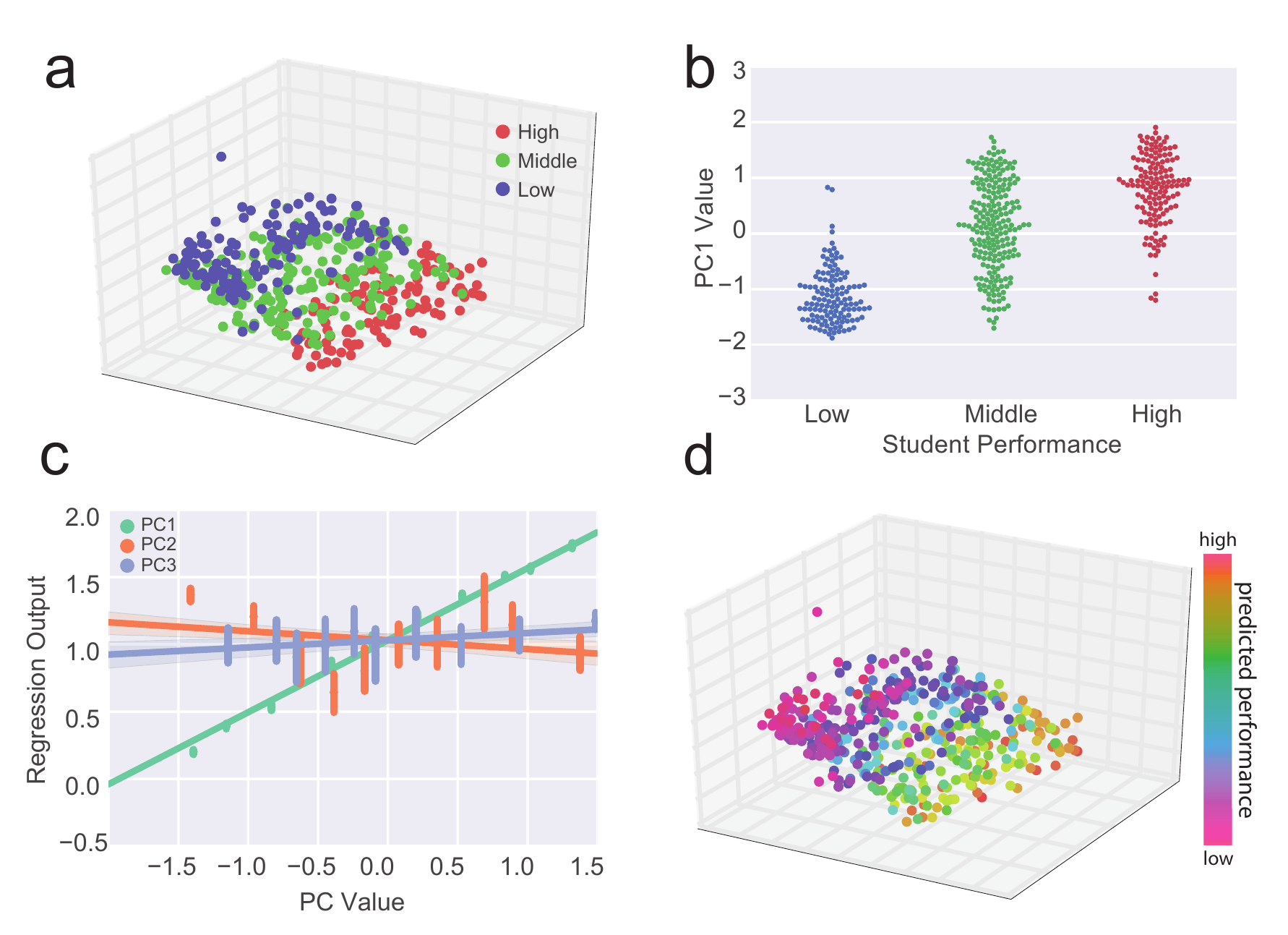}
\caption{\textbf{Relationship between student attributes and performance.} \textbf{a.} Each point represents the feature vector associated with a student, and the points are colored by student performance (red: high, green: middle, blue: low). \textbf{b.} Swarm plot of the first principal component split by student performance (coloring same as above). \textbf{c.} Predicted student performance from linear regression of each PC on student performance.  \textbf{d.} Same as (a), but points are colored by linear regression predictions of student performance by PC1 value (red to violet gradient represents high to low predicted performance).}
\label{fig:education}
\end{figure}

In contrast to the mushroom dataset, where the samples formed clear clusters, the distribution of samples in this dataset appear to form a single contiguous mass.  Further, coloring each sample (student) by their performance rating reveals a striking correspondence between the student's attributes and performance ratings (Fig.~\ref{fig:education}a).  For example, as the student attributes vary along the first principal component, the student performance ratings appear to transition smoothly from low, to medium, to high (Fig.~\ref{fig:education}b).  To highlight this pattern, we fit a linear regression model whose output variable was student performance and the input variables were the first three principal components (Fig.~\ref{fig:education}c).  In this way, the regression model's outputs provide a continuous estimate of student performance, whereas the original data contained only discrete (categorized) estimates.  In Figure~\ref{fig:education}d, each dot from Panel~a has been re-colored according to the regression model's performance predictions, resulting in a smooth gradient from low to high performance.

\subsection*{Example \stepcounter{example}\arabic{example}: Exploring linguistic data from presidential nominees' Twitter posts.}
Whereas the above examples illustrate how simple numerical and categorical features are processed by \hypertools~to reveal geometric patterns in the data, we can use a similar approach to extract and visualize more complex features.  For example, topic models~\cite{BleiEtal03} may be used to derive a vector representation of each document in a corpus according to its linguistic properties.  Specifically, topic models identify ``themes'' that are reflected in varying amounts by different documents in the corpus, where each theme (\textit{topic}) is defined formally as a distribution over words in the vocabulary.  In other words, a neuroscience-themed topic might heavily weight words like \textit{neuron} and \textit{brain}, whereas a sports-themed topic might heavily weight words like \textit{running} and \textit{athlete}.  (Fitting a topic model to a text corpus reveals what the specific topics are and how much each document reflects each topic.)  Once we have derived topic vectors for each document in the corpus, we can use \hypertools~to visualize the full corpus to potentially gain insights into its geometric structure.

As an example of this approach, we next turn to an analysis of Twitter data (``tweets'') from the Twitter accounts of Hillary Clinton (\href{https://twitter.com/HillaryClinton}{@HillaryClinton}) and Donald Trump (\href{https://twitter.com/realDonaldTrump}{@realDonaldTrump}) over the course of their 2016 political campaigns.  The dataset, sourced from \href{https://github.com/fivethirtyeight}{FiveThirtyEight}, contains 6,444 tweets sent from the candidates' primary Twitter accounts between April 17, 2016 and September 26, 2016.

We began our analysis by fitting a 20-topic topic model to the entire collection of tweets from both candidates, yielding a single topic vector for each tweet.  Separately for each candidate, we next computed daily average topic vectors over the six month interval covered by the dataset, and we used \hypertools~to visualize the resulting day-by-day Twitter topics.
    
\begin{figure}[tbp]
\centering
\includegraphics[width=1\textwidth]{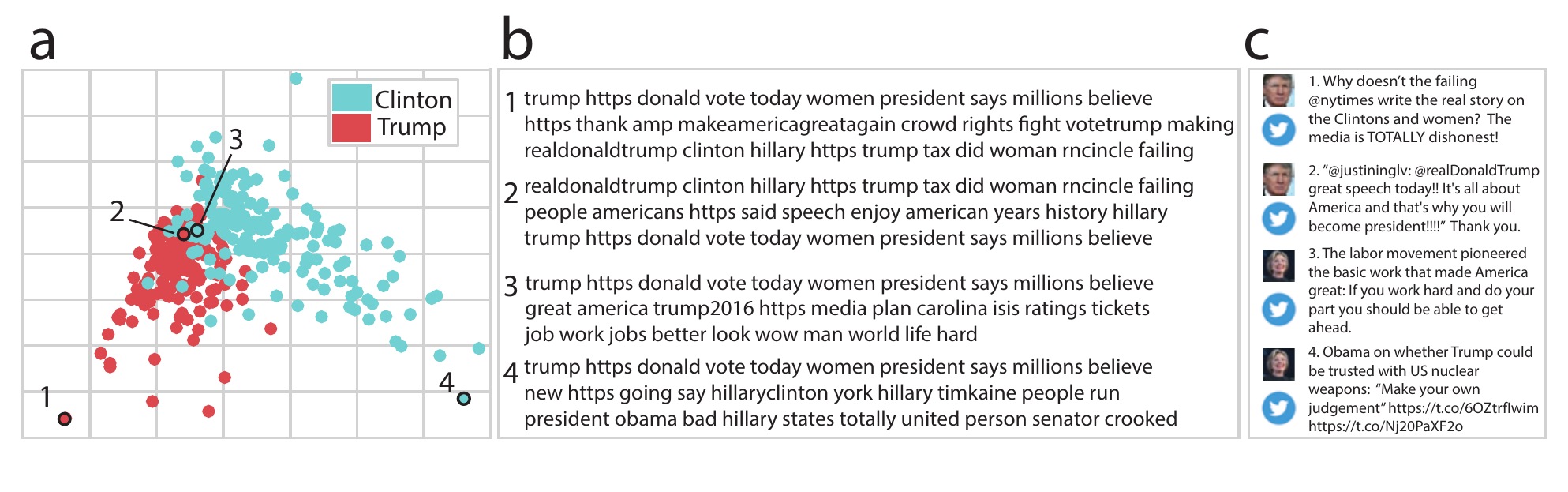}
\caption{\textbf{Topic models of political Twitter data.} \textbf{a.} Two-dimensional representation of Clinton's (blue) and Trump's (red) day-by-day tweet content. \textbf{b.} Top ten words from each of the top three topics on selected days. \textbf{c.} Representative tweets from the selected days.}
\label{fig:tweets}
\end{figure}

Plotting the candidates' tweet content in two dimensions reveals that Clinton's and Trump's tweets were primarily about different topics, resulting in a V-like topic cloud (Fig.~\ref{fig:tweets}a).  We leveraged this structure revealed by \hypertools~to select several days of interest to examine further.  Specifically, we examined (1) a day of Trump tweets whose topic coordinates were especially Trump-like (i.e.\ at the end of the Trump side of the V), (2) a day of Trump tweets whose topic coordinates fell at the intersection of the V, (3) a day of Clinton tweets whose topic coordinates fell at the intersection of the V, and (4) a day of Clinton tweets that fell at the end of the Clinton side of the V.  For example, we wondered whether the candidates' tweets that fell at the extreme ends of the V might be especially representative of each candidates' unique features, whereas tweets that fell at the intersection of the V might express points of similarity between the candidates.  Figure~\ref{fig:tweets}b displays the top 10 words from each of the top three topics for each of these days of interest, and Figure~\ref{fig:tweets}c provides representative tweets from each day.  Strikingly (perhaps), the most Trump-like tweets appear to disparage Clinton, the most Clinton-like tweets appear to disparage Trump, and the overlapping tweets appear to praise America's greatness.

\subsection*{Example \stepcounter{example}\arabic{example}: Cyclical increases in global temperatures over time.}
In addition to generating static point cloud plots, \hypertools~may be used to generate trajectory plots to illustrate dynamic patterns in the data.  To highlight this feature, we used a global temperatures dataset which we acquired from \href{http://berkeleyearth.lbl.gov/city-list/}{Berkeley Earth}.  The Berkeley Earth averaging method takes temperature observations from a large array of weather monitoring stations throughout the world and produces a time-varying estimate of the underlying global temperature field across all of the Earth's land areas. This temperature field may then be sampled to obtain location-specific temperature estimates.

\begin{figure}[tbp]
\centering
\includegraphics[width=1\textwidth]{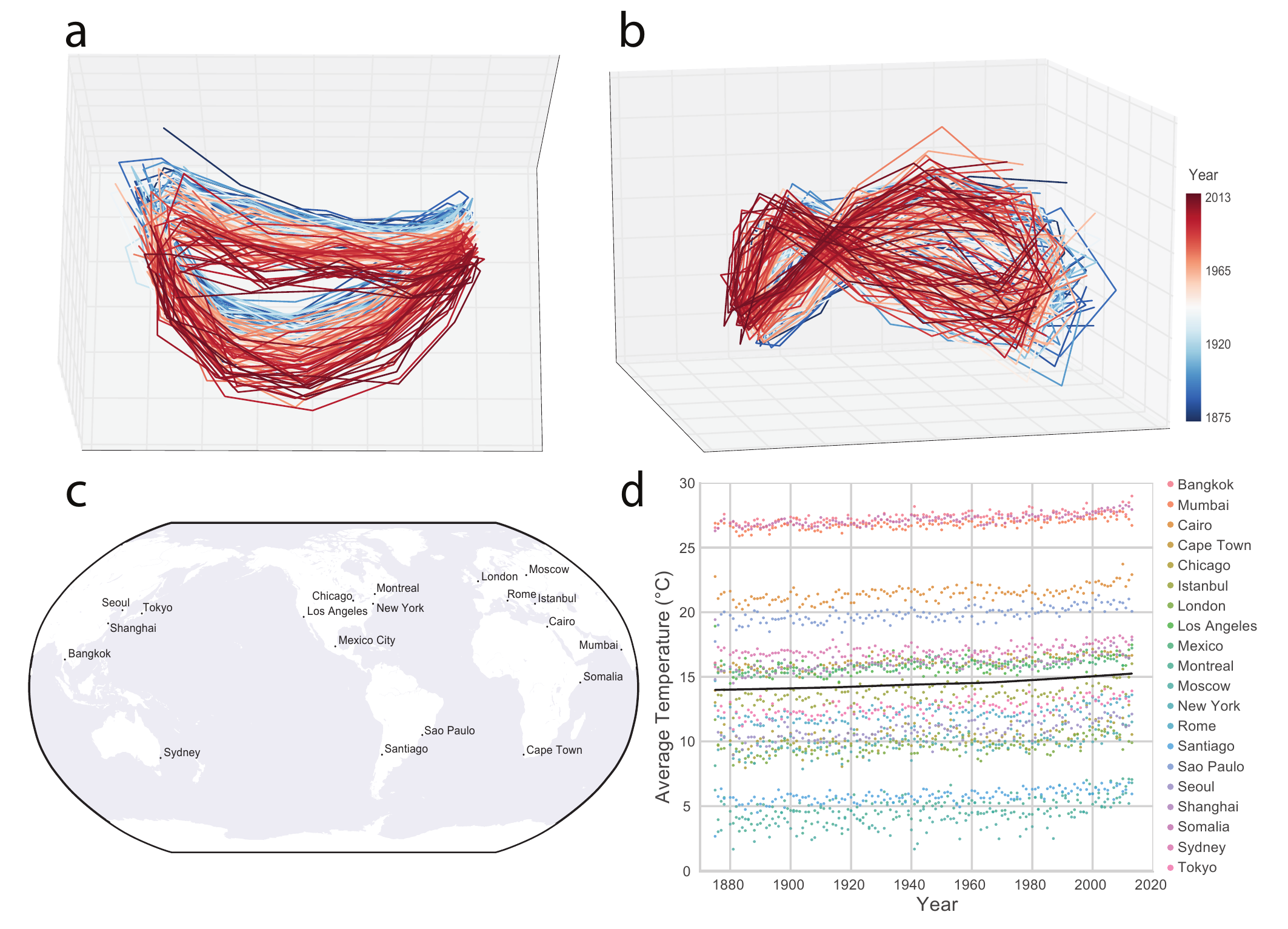}
\caption{\textbf{Global temperatures from 1875--2013.}  \textbf{a.} and \textbf{b.} The global temperatures dataset plotted using PCA dimensionality reduction in two views. The line colors change over time (from the earliest time point in blue to the most recent time point in red).  The view on the left shows the temporal progression in one of the dimensions while the view on right highlights the cyclical nature of the dataset. \textbf{c.} Locations of the 20 cities in the dataset. \textbf{d.} Yearly mean temperatures colored by location and black LOWESS line fit to the full dataset.} \label{fig:globalwarming}
\end{figure}

To visualize how the global temperature field changes over time, we acquired monthly average temperature estimates for 20 cities throughout the world (Fig.~\ref{fig:globalwarming}c) over the 138 year interval from 1875--2013.  We used \hypertools~to plot the resulting temperature trajectory (Fig.~\ref{fig:globalwarming}a,b).  To visualize systematic changes over time, we plotted the month-by-month trajectory for each year in a different color using the \texttt{group} keyword argument to \texttt{plot}:
\begin{align}
\mathtt{hyp.plot(data, group=years, palette='RdBu\_r')}
\end{align}
Two general trends were revealed by plotting the temperature data in this way.  First, the month-by-month temperatures within a year are cyclical (e.g.\ reflecting the changing seasons), which appears in the trajectory as a ``figure 8'' (this trend is most visible in Fig.~\ref{fig:globalwarming}b).
Second, there has been a systematic shift in global temperatures over the 138 year period we examined.  This appears as a systematic shift in the position of the trajectory over time (Fig.~\ref{fig:globalwarming}a), and can also be seen by directly plotting the temperatures over time (Fig.~\ref{fig:globalwarming}d).

\subsection*{Example \stepcounter{example}\arabic{example}: Visualizing the correspondence between neural trajectories and a movie stimulus}
In addition to providing plotting tools for visualizing complex data, \hypertools~also provides tools for aligning trajectories from different sources (see \textit{Align}).  For example, suppose we have brain recordings from different people who all watched the same movie.  The general shapes of different people's brain data trajectories (showing how everyone's brain responses changed over time while watching the movie), as well as the movie trajectory (showing how the movie itself changed over time), might all share similar properties (e.g.\ reflecting the covariance structure of the movie and how people responded to it).  However, different people's brains may have reflected those similar responses differently, and the dimensions of ``brain space'' and ``movie space'' are not directly comparable.  As described in \textit{Materials and Methods}, the \hypertools~toolbox provides an easy-to-use interface for aligning datasets.  In this section example we demonstrate some uses of the \texttt{align} function using a previously published fMRI dataset~\cite{HaxbEtal11}, available for download \href{https://github.com/HaxbyLab/raiders\_data}{here}. The dataset comprises voxel responses from ventral temporal cortex, from each of 11 people, as they watched the feature-length film \textit{Raiders of the Lost Ark}.  The data were processed and hyperaligned as described in the original manuscript~\cite{HaxbEtal11}.

\begin{figure}[tbp]
\centering
\includegraphics[width=\textwidth]{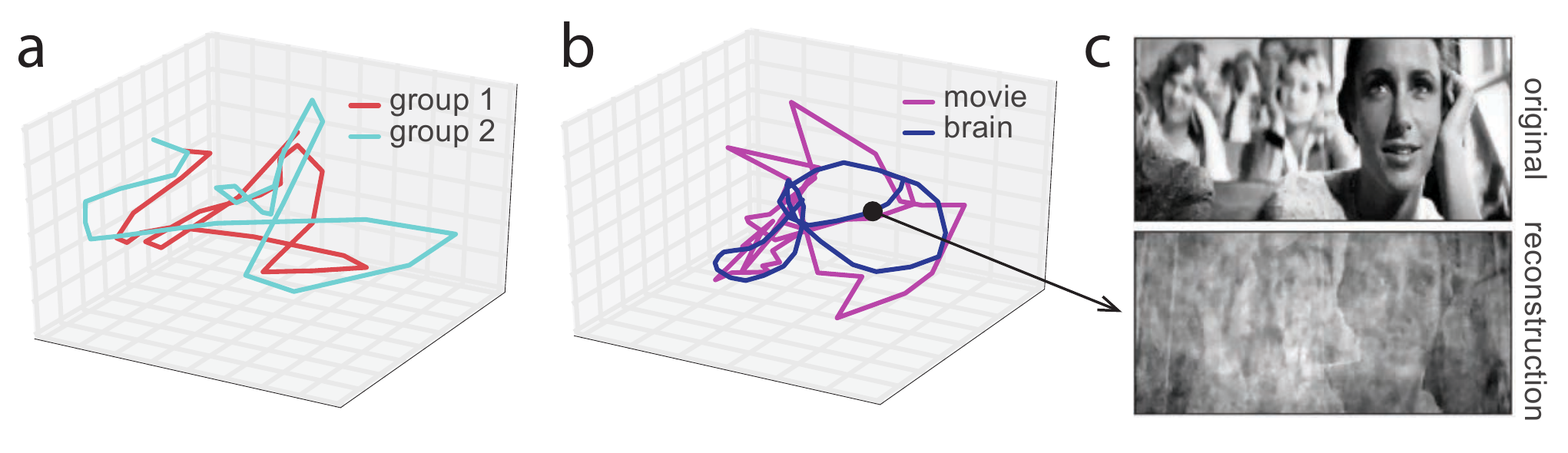}
\caption{\textbf{Brain/movie trajectories during movie viewing.} \textbf{a.} Group-averaged trajectory of brain activity from ventral visual cortex split into two randomly-selected groups of subjects (group 1: $n=6$, group 2: $n=5$) watching the same movie. \textbf{b.} Group-averaged trajectory of brain activity from ventral visual cortex and trajectory of movie (pixel intensities over time) hyperaligned to a common space. \textbf{c.} Movie frame reconstructed from ventral visual brain activity that is aligned to movie space.}
\label{fig:raiders}
\end{figure}

Figure~\ref{fig:raiders}a displays the trajectory plots for the averaged hyperaligned brain responses from two groups of participants in the original experiment (six in group 1, the remaining five in group 2).  The trajectories appear similar in their overall shape (indicating that the two groups of participants had roughly similar brain responses to the movie), but the alignment is imperfect (indicating that understanding individual differences between people's responses might be an interesting future direction to explore).

We next demonstrate how \hypertools~may be used to visualize the correspondence between datasets with different coordinate systems-- specifically, time-varying brain responses to the movie and the time-varying pixel intensities of the movie frames.  To align these spaces, we first preprocessed the movie frames to convert the movie into the $S \times F$ matrix format required by \hypertools~(here $S$ is the number of movie frames and $F$ is the number of pixels per frame).  We downsampled the movie frames from $540 \times 960$ RGB pixels at 30 FPS to $108 \times 192$ grayscale pixels at 1~FPS.  We then re-shaped each downsampled frame into a 20,736-dimensional vector.

We next averaged the (hyperaligned) brain responses from the 11 experimental participants to obtain a single brain response matrix.  We used piecewise cubic interpolation~\cite{FritCarl80} to re-sample this averaged brain response matrix from the original data acquisition rate (one image acquired every 2.5~s) to the downsampled movie frame rate (one image per second).  We used the \texttt{reduce} function to project both the movie and brain data onto 6,641 dimensions (i.e.\ the number of voxels in the original brain data) and shifted the time labels of the brain matrix backwards by 5~s to account for the hemodynamic response.  We then used the \texttt{procrustes} function to align the brain and movie data:
\begin{align}
\mathtt{brain\_aligned\_to\_movie = hyp.tools.procrustes(movie\_data, brain\_data)}
\end{align}
The resulting aligned brain data matrix may then be plotted in the same space as the movie data matrix (Fig.~\ref{fig:raiders}b).  This visualization can provide insights into the similarities and differences between the geometric structure of the original movie and the structure of the brain responses to the movie.

In addition to facilitating visual comparisons of the geometries of the movie and brain data, the aligned data may also be compared in the ``native'' data space.  For example, each coordinate of ``movie space'' corresponds to an image, which may be displayed and examined.  Aligning the brain data to this movie space (using the \texttt{procrustes} function) means that each brain pattern now corresponds to a coordinate in movie space, and therefore the corresponding image may also be displayed and examined (Fig.~\ref{fig:raiders}c).  This provides a means of viewing the original movie through the ``lens'' of the brain responses to that movie.  This general approach could also be carried out in a cross-validated way (i.e.\ using one portion of the data to compute the Procrustean transformation from brain space to movie space, and then applying that transformation to the held-out brain data).  We plan to explore this form of alignment-based decoding in future work.

\section*{Discussion}

Visualizing high-dimensional data via low-dimensional embeddings provides an intuitive means of exploring the geometric and statistical properties of complex datasets.  This can help to guide analysis decisions and facilitate hypothesis generation and testing.  Returning briefly to the example of Anscombe's quartet we discussed in the \textit{Introduction} (Fig.~\ref{fig:anscomb}), striking differences between datasets with very different geometries may be overlooked when solely considering their summary statistics, and this principle can be extended to high-dimensional data as well.  Our \hypertools~toolbox aims to assist in high-dimensional data visualization by providing a simple (yet powerful) set of plotting functions and data manipulation tools.

We have provided brief examples of how our toolbox may be used to examine data from a wide array of domains: geometry (Example 1), biological data (Example 2), educational and sociological data (Example 3), political and linguistic data (Example 4), and neuroscientific data (Example 5).  We chose these particular examples to showcase a broad sampling of the types of visualizations and analyses our toolbox supports, but they are not intended to indicate that our toolbox may be used in only these ways or in these domains.

We hope that \hypertools~will prove useful in analyzing and visualizing complex data from a wide array of domains.  We have released the toolbox under an open-source license to facilitate transparency and widespread adoption.  We also hope that users will contribute to the toolbox by providing feedback and suggestions, and by sharing their own extensions and applications with the community.

\section*{Acknowledgments}
We are grateful for useful discussions with Luke J. Chang and Matthijs van der Meer.  We are also grateful for the help of J. Swaroop Guntapalli in implementing our \texttt{align} function. Our work was supported in part by NSF EPSCoR Award Number 1632738.  The content is solely the responsibility of the authors and does not necessarily represent the official views of our supporting organizations.
\nolinenumbers

\bibliography{memlab}

\bibliographystyle{abbrv}

\end{document}